\documentclass[format=sigchi, review=false, screen=true]{acmart}
\renewcommand\footnotetextcopyrightpermission[1]{}
\usepackage{booktabs} % For formal tables
\usepackage{todonotes}
\begin{document}
\title{Automatic Playlist Continuation through a Composition of Collaborative Filters}
%\titlenote{Produces the permission block, and
%  copyright information}
%\subtitle{Extended Abstract}

%\subtitlenote{The full version of the author's guide is available as
%  \texttt{acmart.pdf} document}

\author{Irene Teinemaa}
%\authornote{}
\affiliation{%
  \institution{University of Tartu}
  \streetaddress{J. Liivi 2}
  \city{Tartu}
  \country{Estonia}
}
\email{irene.teinemaa@ut.ee}

\author{Niek Tax}
%\authornote{}
\affiliation{%
  \institution{Eindhoven University of Technology}
  \streetaddress{5612 AZ}
  \city{Eindhoven}
  \country{Netherlands}
}
\email{n.tax@tue.nl}

\author{Carlos Bentes}
% \authornote{}
\affiliation{%
  \institution{STACC}
  \streetaddress{Ulikooli 2}
  \city{Tartu}
  \country{Estonia}}
\email{carlos.bentes@stacc.ee}

\renewcommand{\shortauthors}{I. Teinemaa et al.}

%-------------------------------------------
% Abstract
%-------------------------------------------
\begin{abstract}

The RecSys Challenge 2018 focused on automatic playlist continuation, i.e., the task was to recommend additional music tracks for playlists based on the playlist's title and/or a subset of the tracks that it already contains. The challenge is based on the Spotify Million Playlist Dataset (MPD), containing the tracks and the metadata from one million real-life playlists. This paper describes the automatic playlist continuation solution of \emph{team Latte}, which is based on a composition of collaborative filters that each capture different aspects of a playlist, where the optimal combination of those collaborative filters is determined using a Tree-structured Parzen Estimator (TPE). The solution obtained the 12th place out of 112 participating teams in the final leaderboard. Team Latte participated in the main track of the challenge of the RecSys Challenge 2018.
%\todo{Niek: My general feeling is that the section titles are wrong. Putting part of the approach in a section called "The dataset" and part of the approach in a section called "Experimental results" leads to the situation where our approach section is very thin, while this should be the main purpose of the paper.}

\end{abstract}
\maketitle

%-------------------------------------------
% Introduction
%-------------------------------------------
\section{Introduction}
With the increasing popularity of online music streaming services, the task of selecting relevant and personalized content in large music catalogs becomes important to avoid choice overload \cite{bollen2010understanding}.
An open challenge in the area of personalization for music recommender systems is known as \emph{automatic playlist continuation} (APC), where the task is to recommend tracks that are likely to be selected as additional tracks for an existing playlist. In APC it is important to recommend relevant content while, at the same time, respecting the characteristics of the original playlist \cite{schedl2018current}. For example, the recommended songs for the continuation of a playlist that consists of Christmas songs should be other Christmas songs.

To promote progress in the area of APC, the RecSys Challenge 2018 focuses on this task. The challenge was organized by Spotify, The University of Massachusetts, Amherst, and Johannes Kepler University, Linz, and was open for submissions from January to July 2018.
In the competition, participants were challenged to create a recommendation system for APC using a dataset of one million playlists that have been created by Spotify users in North America.

The competition task was to generate a list of 500 tracks as playlist continuation for each of the 10000 playlists in the challenge dataset. The playlists in the challenge set were divided into ten \emph{challenge categories}, based on the number of seed tracks (the tracks that are already known to be present in the playlist) and the availability of the playlist title.

%, where every playlist in the challenge dataset can be grouped in distinct sets accordingly with its the number of initial tracks (designated as seed tracks).
%In all cases, the model should be able to produce a final playlists with fixed size of 500 tracks (not including the seed playlist tracks or duplicated tracks).\todo{Niek: the groups and the fixed size of the final playlists doesn't seem relevant for an intro. I would move this to Sec 2. It would be better to use this space to already give a 1-paragraph summary of our approach.}

This manuscript describes the solution proposed by Team Latte. The main idea behind the proposed approach is to construct several collaborative filters, based on the co-occurrence of tracks with other tracks, artists, albums, and words in the playlist title. Furthermore, the solution adopts a specialized optimization strategy, where the weights of each collaborative filter are optimized locally within each challenge category using an optimization method called Tree-structured Parzen Estimator (TPE)~\cite{bergstra2011algorithms}. The final recommendation scores are produced as a weighted sum of the individual collaborative filtering components, and then post-processed using several heuristic strategies.

The remainder of this paper is organized as follows: 
Section~\ref{sec:dataset} describes the data provided during the competition. 
%Section~\ref{sec:evaluation} presents the main metrics and strategy used to evaluate model performance.
Section~\ref{sec:model} describes the proposed framework based on several collaborative filters and their combination.
Section~\ref{sec:setup} further details the model optimization and selection procedures to combine the collaborative filers, and discusses some results on internal validation sets.
Finally, in Section~\ref{sec:conclusion} we conclude the paper.

%-------------------------------------------
% Dataset
%-------------------------------------------
\section{Dataset}
\label{sec:dataset}
The dataset consists of one million playlists created by Spotify users and distributed as the Million Playlist Dataset (MPD) for exclusive use in the competition. The MPD dataset includes information about the playlist (title, identification, number of artists, playlist duration) and track information (album name, identification, artist, track duration, track name) for every playlist. A complete description of the dataset can be found in \cite{spotify2018mpd}.
%\todo{Niek: MPL? It's MPD, right (https://labs.spotify.com/2018/05/30/introducing-the-million-playlist-dataset-and-recsys-challenge-2018/)}

In addition to the MPD dataset, the organizers provided the \emph{challenge dataset}, i.e. an official test dataset that contains partial information about 10000 playlists: the playlist title and/or a number of \emph{seed tracks} (a subset of tracks present in the playlist). The aim of the challenge was to generate a list of 500 \emph{recommended tracks} for each of these playlists, based on the partial information available. Additionally, each playlist contained a number of \emph{holdout tracks} that were known only to the organizers of the challenge. The submissions of the participants were evaluated and ranked based on the correspondence between the recommended tracks and the holdout tracks~\cite{spotify2018challengerules}.

The playlists in the challenge dataset can be divided into ten distinct challenge categories based on the type of the provided partial information \cite{spotify2018challengerules}, namely:

\begin{itemize}
\item G1: Playlists with a title only
\item G2: Playlists with a title and the first track
\item G3: Playlists with a title and the first five tracks
\item G4: Playlists with first five tracks (no title)
\item G5: Playlists with a title and the first ten tracks
\item G6: Playlists with first ten tracks (no title)
\item G7: Playlists with a title and the first 25 tracks
\item G8: Playlists with a title and 25 random tracks
\item G9: Playlists with a title and the first 100 tracks
\item G10: Playlists with a title and 100 random tracks
\end{itemize}

\begin{comment}
Playlists in $D_{test}$ are grouped accordingly to the total number of seed tracks $K$, namely:
$K=0$ for empty playlists;
$K=1$ for playlists containing only one track;
$K=5$ for playlists containing five tracks;
$K=10$ for playlists containing ten tracks;
$K=25$ for playlists containing twenty five tracks;
and $K=100$ for playlists containing one hundred tracks;
\end{comment}

Table~\ref{tab:challenge_dataset} shows the distribution of playlists, the average number of holdout tracks ($H_\mathit{avg}$), the number of unique seed tracks ($N_\mathit{Track}$), and the number of unique artists ($N_\mathit{Artist}$) present the challenge dataset, grouped according to the number of seed tracks $K$.

\begin{table}[tb]
  \caption{Challenge Dataset Statistics}
  \label{tab:challenge_dataset}
  \begin{tabular}{ccccc}
    \toprule
    Group & $N_\mathit{Playlist}$ & $H_\mathit{avg}$ & $N_\mathit{Track}$ & $N_\mathit{Artist}$ \\
    \midrule
    K=0		& 1000		& 29	& 0 		& 0 \\
    K=1		& 1000		& 23	& 932 		& 715 \\
    K=5		& 2000		& 55	& 6790 		& 2762 \\
	K=10	& 2000		& 53	& 11877 	& 4096 \\
	K=25	& 2000		& 126	& 22507 	& 6253 \\
    K=100	& 2000		& 88	& 53552 	& 11517 \\
  \bottomrule
\end{tabular}
\end{table}

%-------------------------------------------
% Model
%-------------------------------------------
\section{Framework}
\label{sec:model}
In this section, we describe the framework for automatic playlist continuation (APC). The framework consists of three steps: first a collection of multiple different collaborative filtering models are extracted in the \emph{collaborative filtering stage}, then, the predictions of the collaborative filtering models are combined into a single relevance prediction per playlist-track combination in the \emph{composition stage}, and finally, the recommendations are generated in the \emph{playlist continuation stage}. We now continue with describing these stages in detail.
%\todo{Niek: This section seems very thin, while it seems from the CfP that the approach section is the main thing that they want to see in the paper. I would add a lot more detail. For example, the NLP part to construct the matrices is completely missing, and so is the description of our tricks to keep the matrix sizes under control.}
\subsection{Collaborative Filtering Stage}

The task of collaborative filtering is to predict the utility of items (tracks) to a particular \emph{context} (playlist) based on vector similarities between these entities extracted from data \cite{breese1998empirical}. This \emph{context} can be based on different aspects of a playlist. For example, in item-item collaborative filtering, the context is based on the tracks that are already present in the playlist.
A total of four collaborative models were built in order to capture different contexts:
\begin{description}
\item[track-track model ($M_u$)]{models the relevance of a given track for a given playlist based on the set of tracks that are currently present in the playlist. This model is a traditional item-item collaborative filtering model.}
\item[word-track model ($M_w$)]{models the relevance of a given track for a given playlist based on the name of the playlist. This collaborative filter that models the relation between words in the playlist name and the occurrence of tracks when a playlist contains this word in the playlist title. The words are extracted from the playlist names by splitting the playlist name on the space character (i.e. ' '), transforming the results to lowercase, and removing punctuation marks.}
\item[album-track model ($M_\mathit{al}$)]{models the relevance of a given track for a given playlist based on the albums from the tracks that are currently present in the playlist. This collaborative filter models the relation between the set of albums of the tracks that are currently in the playlist and the occurrence of tracks when these albums are in the playlist.}
\item[artist-track model ($M_\mathit{ar}$)]{models the relevance of a given track for a given playlist based on the artists the created the tracks that are currently present in the playlist. This collaborative filter models the relation between the set of albums of the tracks that are currently in the playlist and the occurrence of tracks when these albums are in the playlist.}
\end{description}

\subsection{Composition Stage}

The output of every collaborative filter is combined in a final ranking model ($M_{c}$) using a weighted sum given by: 

\begin{equation}
  M_{c} = W_u * M_u + W_w * M_w + W_\mathit{al} * M_\mathit{al} + W_\mathit{ar} * M_\mathit{ar}
\end{equation}

where $W_u$, $W_w$, $W_\mathit{al}$ and $W_\mathit{ar}$ are real-valued weights in range $[0,1]$.

The best configuration of weights is found using an optimization procedure, such as Tree-structured Parzen Estimator (TPE)~\cite{bergstra2011algorithms}. We experiment with two types of weighting schemes: 1) global weights (optimized over all instances) and 2) local weights (optimized separately for each challenge category). We describe the procedure to determine the weights $W_u$, $W_w$, $W_\mathit{al}$, and $W_\mathit{ar}$ in detail in Section~\ref{sec:setup}.

\subsection{Playlist Continuation Stage}
To determine the recommended tracks for a given playlist, we filter the tracks on $M_c>0$ and then sort the tracks in descending order based on their $M_c$ value, using the $M_c$ value that uses the weights that we found in the composition stage. However, it can be the case that fewer than 500 tracks have a value of $M_c$ that is larger than zero, in which case the requirement of recommending 500 songs would not be satisfied.
To improve the order of tracks in the recommendations ranking and to guarantee a total 500 recommended tracks for every playlist, we apply two post-processing steps.

The first post-processing step aims at completing the albums that are currently already present in the playlists. This is motivated by the fact that a reasonable number of playlists in the dataset contained exactly all the tracks of a single album, and we found the $M_c$ to be insufficient to properly detect this scenario and complete the album for playlists that contain a high number of tracks from the same album. When the ratio of the number of tracks from the number of distinct albums that are currently in the playlist exceeds a threshold $m$ (where $m$ is a tunable parameter), we first recommend all the tracks from that remaining album before recommending the tracks based on $M_c$.

%\subsection{Complementing with popularity model}
As a second post-processing step, to fulfill the requirement of recommending exactly 500 tracks, we append the list of recommended tracks with the most popular tracks in the dataset in decreasing order of overall frequency until the list of recommended tracks contains exactly 500 tracks.

%-------------------------------------------
% Results
%-------------------------------------------
\section{Model selection}
\label{sec:setup}
%\todo{Niek: Again this section seems to have the wrong title, as the model optimization is part of approach. I would simply call this section "Optimizing Model Weight" or something like that.}
We evaluate the model instantiations using a combination of three measures R-precision, NDCG, and CLICKS, which are the same three measures that are used by the RecSys challenge organizers to score the submissions. We select the best performing model instantiation for submission. 

In this section we present the evaluation measures, the procedure for optimizing the models' parameters, the procedure and the results for selecting the best model. The framework was implemented in Python and can be found at \cite{latte2018recsys} under open source license.

\subsection{Evaluation measures}
\label{sec:evaluation}

The R-precision, defined as:

\begin{equation}
  \text{R-precision} = \frac{|G \cap R|}{|G|}
\end{equation}

where $G$ is the set of ground truth (holdout) tracks, and $R$ is the set of recommended tracks. The notation $|.|$ denotes the number of elements in the set.

The Normalized discounted cumulative gain (NDCG), defined as:

\begin{equation}
  \label{equ:ndcg}
  \text{NDCG} = \frac{DCG}{IDCG}
\end{equation}

where:

\begin{equation}
  \text{DCG} = rel_1 + \sum_{i=2}^{|R|} \frac{rel_i}{log_2(i+1)}
\end{equation}

\begin{equation}
  \text{IDCG} = 1 + \sum_{i=2}^{|G|} \frac{1}{log_2(i+1)}
\end{equation}

The Recommended Songs CLICKS metric, that mimics a Spotify feature for track recommendation where ten tracks are presented at a certain time to the user as the suggestion to complete the playlist. This metric captures the number of refreshes needed before a relevant track is encountered, and is defined as:

\begin{equation}
  \text{CLICKS} = \frac{arg min_i \{R_i:R_i \in G\} - 1}{10}
\end{equation}

While NDCG and CLICKS were calculated based on the track-level agreement between the holdout tracks and the recommended tracks, R-precision was calculated on the artist-level agreement. In other words, it was considered sufficient if the artist of a recommended track matched the artist of a holdout track.

\subsection{Approaches}
We tested three instantiations of the proposed framework, namely:
\begin{itemize}
\item composition via global weights;
\item composition via local weights, without album completion (i.e., $m=\infty$);
\item composition via local weights, where the album completion threshold $m$ is optimized through the same procedure as optimizing the weights.
\end{itemize}

As a baseline, we compared the results to a simple popularity-based model, where the recommendation list is created based on the overall popularity of songs in a non-personalized manner.

\subsection{Model Optimization}
For each of the tested approaches, the weights for combining the collaborative filters needed to be optimized. Furthermore, in the variant with album completion, the song to album ratio $m$ was optimized. To this end, we extracted a optimization dataset ($D_\mathit{opt}$) containing 10k playlists (playlists $980001-990000$ from the MPD). Similarly to the original challenge dataset, we divided these playlists into 10 distinct categories that match the challenge categories (see Section~\ref{sec:dataset}) via random sampling. The statistics of the $D_\mathit{opt}$ dataset can be seen in Table~\ref{tab:opt_dataset}.

\begin{table}[tb]
  \caption{Optimization Dataset Statistics}
  \label{tab:opt_dataset}
  \begin{tabular}{lrrrr}
    \toprule
    Group & $N_\mathit{Playlist}$ & $H_\mathit{avg}$ & $N_\mathit{Track}$ & $N_\mathit{Artist}$ \\
    \midrule
    K=0		& 1000		& 38	& 0 		& 0 \\
    K=1		& 1000		& 37	& 942	 	& 758 \\
    K=5		& 2000		& 33	& 7548 		& 3496 \\
	K=10	& 2000		& 33	& 13487 	& 5127 \\
	K=25	& 2000		& 32	& 27185 	& 8789 \\
    K=100	& 2000		& 53	& 76648 	& 18242 \\
  \bottomrule
\end{tabular}
\end{table}

The optimization process was set maximize the NDCG metric (Equation~\ref{equ:ndcg}) and was executed using Tree-structured Parzen Estimator (TPE)~\cite{bergstra2011algorithms}, which is a type of a Sequential Model-Based Global Optimization (SMBO)~\cite{Hutter2011} algorithm. We use the TPE implementation that is available in the Python library Hyperopt~\cite{bergstra2013hyperopt}.
The TPE optimization process was set to run for 100 iterations and the search space of weights defined as a uniform random variable ranging from 0 to 1.

Table~\ref{tab:optimal_weights} shows the optimized best sets of weights separately for each category (used in the local weights composition) and global weights (used in the global weights composition). 

\begin{table}[!htbp] \centering 
  \caption{Optimized weights} 
  \label{tab:optimal_weights} 
\begin{tabular}{lrrrrr} 
\toprule 
Category & $W_u$ & $W_w$ & $W_\mathit{al}$ & $W_\mathit{ar}$ & $m$ \\ 
\midrule
title\_only & $0.000$ & $1.000$ & $0.000$ & $0.000$ & - \\ 
1\_with\_title & $1.000$ & $0.423$ & $0.001$ & $0.011$ & 1 \\ 
5\_no\_title & $1.000$ & $0.000$ & $0.040$ & $0.040$ & 2 \\ 
5\_with\_title & $1.000$ & $0.337$ & $0.006$ & $0.010$ & 2 \\ 
10\_no\_title & $1.000$ & $0.000$ & $0.003$ & $0.009$ & 2 \\ 
10\_with\_title & $1.000$ & $0.964$ & $0.002$ & $0.001$ & 2 \\ 
25\_first & $1.000$ & $0.795$ & $0.001$ & $0.037$ & 2 \\ 
25\_random & $1.000$ & $0.437$ & $0.145$ & $0.022$ & 2 \\ 
100\_first & $1.000$ & $0.828$ & $0.028$ & $0.045$ & 2 \\ 
100\_random & $0.915$ & $1.000$ & $0.169$ & $0.133$ & 3 \\ 
global & $1.000$ & $0.517$ & $0.084$ & $0.056$ & - \\ 
\bottomrule 
\end{tabular} 
\end{table} 

\subsection{Model Selection}
In order to select the best model from the proposed framework in an offline manner (without making an official submission), we extracted a validation dataset ($D_\mathit{val}$) containing 10k playlists (playlists $990001-1000000$ from the MPD). Again, we divided these playlists into ten distinct challenge categories via random sampling. The statistics of the $D_\mathit{val}$ dataset can be seen in Table~\ref{tab:val_dataset}. The validation set was used as a proxy to the challenge leaderboard, guiding model selection and improvements.

%We divide the initial MPD dataset into three distinct disjoint subsets: an optimization dataset ($D_\mathit{opt}$) 
%\todo{Niek: Please use mathit when writing text in math mode to obtain proper spacing between letters} 
%with 10k, a validation dataset ($D_\mathit{val}$) with 10k playlists, and a training dataset ($D_\mathit{train}$) consisting of the 980k remaining playlists. 

%The playlists in $D_\mathit{val}$ and are selected to match the distribution of $D_\mathit{test}$ in terms of number of playlists per group, as shown in Table~\ref{tab:val_dataset} and Table~\ref{tab:opt_dataset}.
%\todo{Niek: This section seems to have the wrong title, as it does much more than just describing the data. Instead, it is offers some exploratory analysis and even partly goes into approach by mentioning the split in train/optimization/validation sets. Maybe it should be called "Exploratory Analysis \& Preprocessing"?}\todo{Niek: we can't introduce "optimization dataset" without first mentioning that we use optimization in our approach. The new paragraph in Sec 1 could mitigate this problem.}

\begin{table}[tb]
  \caption{Validation Dataset Statistics}
  \label{tab:val_dataset}
  \begin{tabular}{lrrrr}
    \toprule
    Group & $N_\mathit{Playlist}$ & $H_\mathit{avg}$ & $N_\mathit{Track}$ & $N_\mathit{Artist}$ \\
    \midrule
    K=0		& 1000		& 38	& 0 		& 0 \\
    K=1		& 1000		& 38	& 943	 	& 737 \\
    K=5		& 2000		& 34	& 7451 		& 3427 \\
	K=10	& 2000		& 33	& 13455 	& 5197 \\
	K=25	& 2000		& 33	& 26864 	& 8512 \\
    K=100	& 2000		& 53	& 80171 	& 18570 \\
  \bottomrule
\end{tabular}
\end{table}

%\subsection{Model Optimization}

%The weights are calculated to every challenge group (see section~\ref{sec:dataset}), in a total of 10 variants per model (total of 40 weights). 

%-------------------------------------------
% Results
%-------------------------------------------
\subsection{Results}
\label{sec:results}

This subsection presents and discusses the results of our experiments.

%\subsubsection{Comparison of Models}

\begin{figure}[tbh]
\centering
\includegraphics[width=1\linewidth]{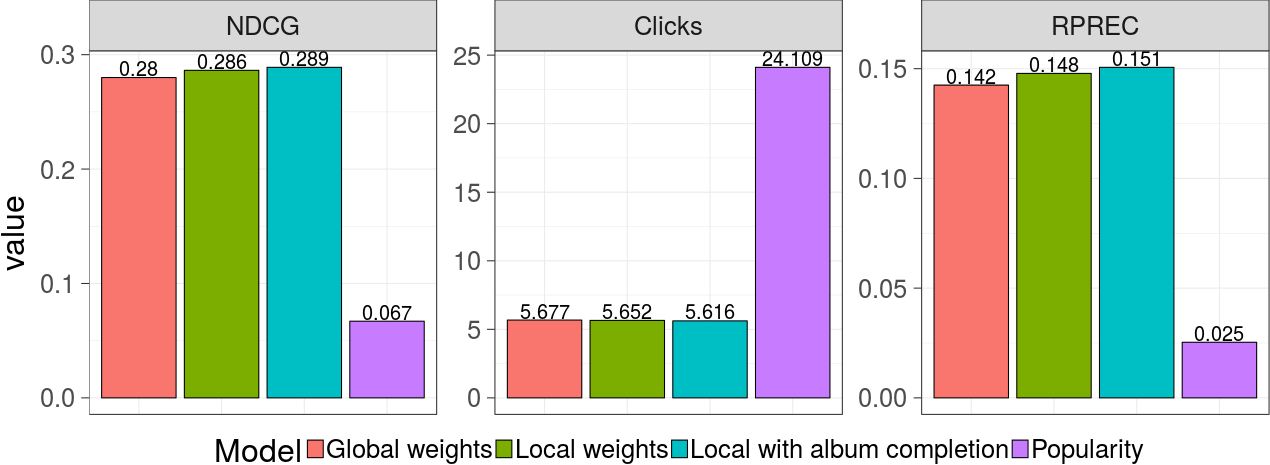}
\caption{Performance of different models on validation set}
\label{fig:models_performance}
\end{figure}

Figure~\ref{fig:models_performance} shows the performance (in terms of NDCG, CLICKS, and RPREC) for the tested instantiations of the framework and the baseline popularity model. Note that in all cases, the composed collaborative model performs better than the popularity model. The model with local weights and album completion is the best performing model and was the selected strategy for our final submission.

%\subsubsection{Final Model}

%As found above, the final model is a combination of collaborative filters with local weights optimized separately for each challenge category with an additional album completion element. 
The results of the final model on both the validation set ($D_\mathit{val}$) and the challenge set are presented in Table~\ref{tab:leaderboard}. 
%The validation set uses 10k playlists not seen during training or optimization.
The Leaderboard score is the score given by the submission website, calculated by the organizers based on the recommended tracks and the holdout tracks (the ground truth values not available to participants) in the challenge dataset.

\begin{table}[tbh]
  \caption{Results of Composed Model}
  \label{tab:leaderboard}
  \begin{tabular}{lrr}
    \toprule
    Metric & Validation & Leaderboard\\
    \midrule
    RPREC	& 0.150587	& 0.203652 \\
    NDCG	& 0.288921	& 0.361175 \\
    CLICKS	& 5.6156	& 2.0240 \\
  \bottomrule
\end{tabular}
\end{table}

To further analyse the performance of the final model within different challenge categories, Table~\ref{tab:results_testgroup} presents the results for the composed model in each of these categories in the validation set\footnote{Note that the overall scores are slightly different than in the above, since this detailed evaluation was executed with training on 400k playlists and a total of 100k tracks only, to reduce the computations.}. We can see in this table that the model is doing considerably better in the groups where the seed tracks were selected randomly from the playlist. The performance is lowest in the category where only the playlist title was provided as input.

\begin{table}[tbh] \centering 
  \caption{Results by challenge category (model trained on 400k playlists, 100k tracks)} 
  \label{tab:results_testgroup} 
\begin{tabular}{lrrrr} 
\toprule 
Category & NDCG & CLICKS & RPREC \\ 
\midrule
title\_only & $0.179$ & $13.990$ & $0.092$ \\
1\_with\_title & $0.266$ & $7.252$ & $0.141$ \\ 
5\_no\_title & $0.270$ & $6.373$ & $0.135$ \\ 
5\_with\_title & $0.271$ & $5.943$ & $0.135$ \\ 
10\_no\_title & $0.259$ & $6.740$ & $0.128$ \\ 
10\_with\_title & $0.263$ & $6.565$ & $0.130$ \\ 
25\_first & $0.258$ & $6.353$ & $0.122$ \\ 
25\_random & $0.353$ & $3.307$ & $0.199$ \\ 
100\_first & $0.219$ & $6.093$ & $0.108$ \\ 
100\_random & $0.372$ & $2.782$ & $0.223$ \\
\midrule
Overall & $0.271$ & $6.540$ & $0.141$ \\ 
\bottomrule
\end{tabular} 
\end{table}

%-------------------------------------------
% Conclusion
%-------------------------------------------
\section{Conclusion}
\label{sec:conclusion}

In the 2018 RecSys challenge, teams competed in the task of automatic playlist competition. To simulate different challenges in the playlist completion task, a challenge dataset was provided with ten different types of seed information (called challenge categories). Our solution was based on combining multiple different collaborative filters that each capture different aspects of a playlist, and we combined them using a Tree-structured Parzen Estimator optimization approach where we optimized the weights locally for each of the challenge categories. The solution strategy shows promising results, ranking our team in position 12 out of 112 teams in the final competition leaderboard.

%-------------------------------------------
% References
%-------------------------------------------
\bibliographystyle{ACM-Reference-Format}
\bibliography{bibliography.bib}

\end{document}